\let\Xdocument\document
\documentclass{iau}
\let\document\Xdocument

\usepackage{amsmath}
\usepackage{graphicx}
\usepackage{multirow}

\begin{document}

\lefttitle{The Local Value of H$_0$}
\righttitle{The Local Value of H$_0$}

\jnlPage{1}{7}
\jnlDoiYr{2021}
\doival{10.1017/xxxxx}

\aopheadtitle{Proceedings of IAU Symposium 376}
\editors{Richard de Grijs, Patricia A. Whitelock and Marcio Catelan, eds.}

\title{The Local Value of H$_0$}

\author{Adam G. Riess and Louise Breuval}
\affiliation{Department of Physics and Astronomy, Johns Hopkins University, Baltimore, MD 21218, USA}
\affiliation{Space Telescope Science Institute, 3700 San Martin Drive, Baltimore, MD 21218, USA}

\begin{abstract}
We review the local determination of the Hubble constant, H$_0$,
focusing on recent measurements of a distance ladder constructed from
geometry, Cepheid variables and Type Ia supernovae (SNe Ia). We
explain in some detail the components of the ladder: (1) geometry from
Milky Way parallaxes, masers in NGC 4258 and detached eclipsing
binaries in the Large Magellanic Cloud; (2) measurements of Cepheids
with the {\sl Hubble Space Telescope} ({\sl HST}) in these anchors and
in the hosts of 42 SNe Ia; and (3) SNe Ia in the Hubble flow. Great
attention to negating systematic uncertainties through the use of
differential measurements is reviewed. A wide array of tests are
discussed. The measurements provide a strong indication of a
discrepancy between the local measure of H$_0$ and its value predicted
by $\Lambda$CDM theory, calibrated by the cosmic
microwave background ({\sl Planck}), a decade-long challenge known as
the `Hubble Tension'. We present new measurements with the {\sl James
Webb Space Telescope} of $>$320 Cepheids on both rungs of the distance
ladder, in a SN Ia host and the geometric calibrator NGC 4258, showing
reduced noise and good agreement with the same as measured with {\it HST}. This provides
strong evidence that systematic errors in {\it HST} Cepheid photometry
do not play a significant role in the present Hubble Tension. Future
measurements are expected to refine the local determination of the
Hubble constant.
\end{abstract}


\maketitle

\section{Why care about the Hubble Constant?}

The Hubble constant, H$_0$, is a measure of the present expansion rate
of the Universe and may be used to infer related quantities such as
the age of the Universe, its fate, the distance to galaxies (with
known redshifts), etc. Connecting the measured value of the Hubble
constant to related properties of the Universe tests our fundamental
understanding of our cosmic paradigm.
  
In the local Universe, $z\sim 0$, $H_0$ is given by the constant of
proportionalty between distances, $D$, and redshifts, $z$, in the
relation $cz={\rm H}_0 D$. More generally, i.e., at $z>0$ and $t<t_0$
(where $t_0$ refers to the present era) the relation between distance
and redshift may evolve and so is more generally defined as
\begin{eqnarray}
D = {cz\over {\rm H}_0}
\Bigg\{ 1 
- 
\left[1+{q_0\over2}\right] {z} 
+
\left[ 1 + q_0 + {q_0^2\over2} - {j_0\over6}   \right] z^2 
+ O(z^3) \Bigg\},
\nonumber
\label{E:physical}
\end{eqnarray}
which follows from a Taylor expansion of the increasing scale factor,
$a$:
\begin{eqnarray}
a(t)= a_0 \;
\Bigg\{ 1 + {\rm H}_0 \; (t-t_0) - {1\over2} \; q_0 \; {\rm H}_0^2 \;(t-t_0)^2 
+{1\over3!}\;  j_0\; {\rm H}_0^3 \;(t-t_0)^3 
\nonumber
+ O([t-t_0]^4) \Bigg\},
\end{eqnarray}
with $H(t) = + ({\rm d}a/{\rm d}t)/a$ the expansion rate at time $t$,
$q(t) = -({\rm d}^2 a/ {\rm d}t^2) \left[ H(t) \right]^{-2}/a$ the
deceleration parameter, and $j(t) =({\rm d}^3 a/ {\rm d}t^3) \left[
H(t) \right]^{-3}/a$ the jerk parameter, and so on. As provided, these
relations ignore curvature but terms for a non-flat Universe can be
included. Thus seen, the Hubble `constant' is the present value of the
expansion H$_0= H(t_0)$ today. The higher-order derivatives of
expansion, $q_0$ and $j_0$, can be directly determined from this
relation, provided a set of distance and redshifts measurements
covering a wide range of redshift is available (e.g., with Type Ia
supernovae, SNe Ia, at $0 < z < 1$ free of cosmological model
assumptions). This approach, following the definition, is called the `Direct'
route, because it does not invoke a cosmological model and has
negligible dependence on gravity.

\begin{figure}[h] 
\begin{center}
\includegraphics[width=0.98\textwidth]{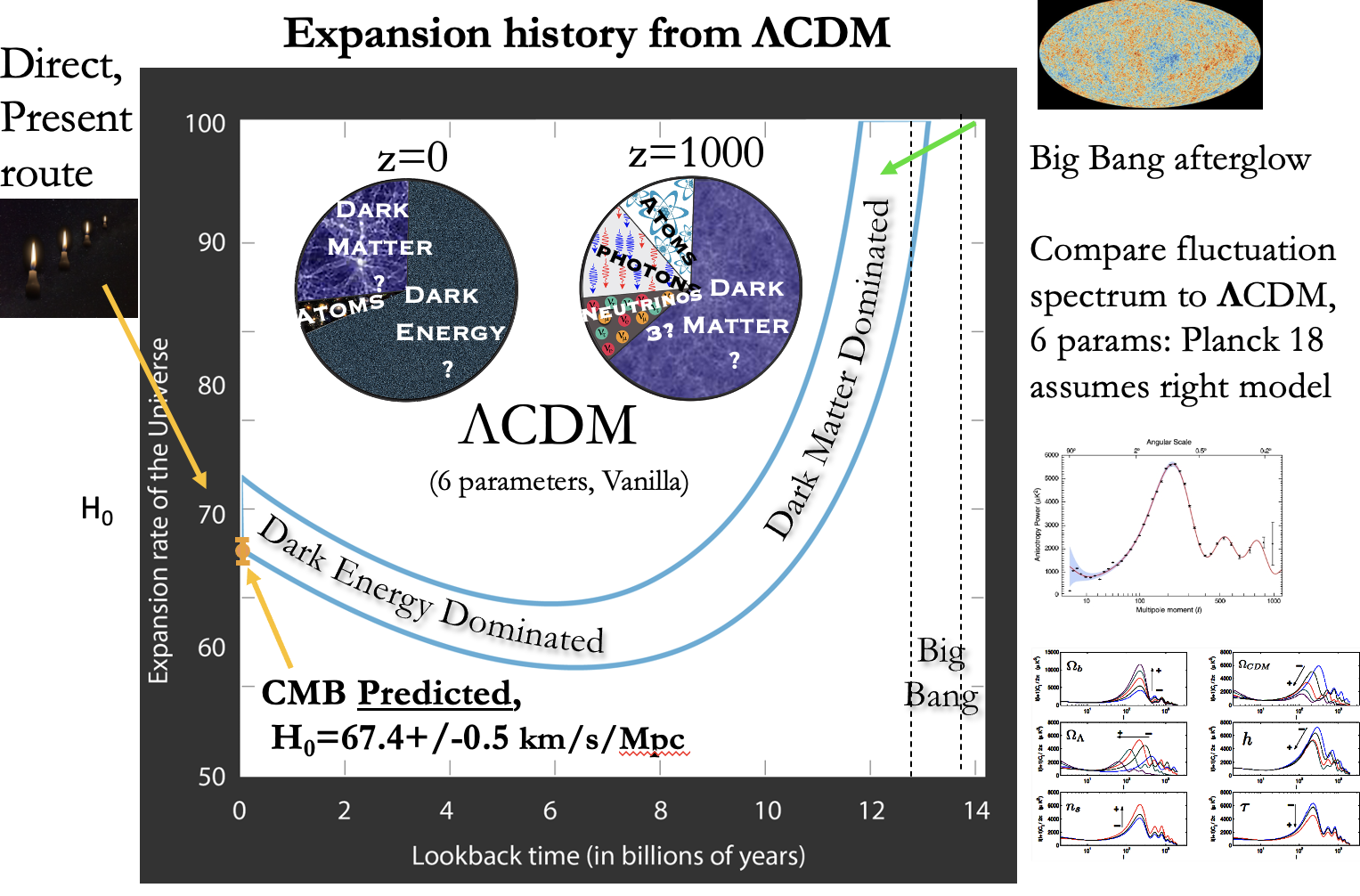}
\end{center}
\caption{The expansion rate of the Universe can be predicted by the
$\Lambda$CDM model with its parameters calibrated by the CMB or
measured directly and locally from redshifts and distances.}
\label{fg:expansionhistory} 
\end{figure}

The alternative route to knowledge of H$_0$ comes from predicting its
present value from fine calibration of the cosmological model,
(`Vanilla') $\Lambda$ cold dark matter ($\Lambda$CDM), measured from
the cosmic microwave background (CMB). By invoking $\Lambda$CDM we are
referring to a description of the Universe as composed of the simplest
dark matter (i.e., non-interacting, non-decaying, from a particle with
only gravitational interactions), the simplest dark energy (i.e., a
`Cosmological Constant'), atoms, photons, and neutrinos (three
species), without spatial curvature, and {\it without any other
cosmologically important features} (hence `Vanilla'). We can use the model as it would
have looked at $z>1000$ to predict the physical size of fluctuations
in the primordial plasma and compare the fluctuation spectrum with their
angular size as observed from the CMB. This comparison serves to
calibrate the six free parameters in $\Lambda$CDM. Once calibrated,
the model predicts that dramatic changes in the Universe will occur
(matter dominated followed by vacuum energy dominated) and describes
the expansion history, $H(z)$, from $z=1000$ to $z=0$, and hence the
value of H$_0$. In the appendix we provide a detailed description of
how, in practice, the value of H$_0$ is measured from the CMB. We
encourage the reader to review it, as it may challenge the prior
belief that this is simple! See \cite{Kamionkowski:2023}
and \cite{Planck:2018} for details. As should be clear, $\Lambda$CDM
is a {\it phenomenological} model with parameters that stand in for a
physical description of the unknown nature of 96\% of the present
Universe. Thus, H$_0$ is uniquely suited to provide an `end-to-end'
test of $\Lambda$CDM and our understanding of the Universe. By
comparing the direct and model-dependent routes, we can test the
model. Figure \ref{fg:expansionhistory} illustrates this test.

\section{The Distance Ladder: From Geometry to Cepheids to Type Ia Supernovae}

Because the indirect route offers better than percent-level precision,
fully leveraging the comparison demands a percent-level local
measurement. The first generation of such measurements from the {\sl
Hubble Space Telescope} ({\sl HST}) Key Project \citep[KP; see][in
this conference]{Freedman:2023}, foundational work of immense importance,
reached 10\% precision by 2001 \citep[see also][]{Sandage:2006}.  
To reach greater precision requires a considerable redesign of the
approach and methods to reach percent-level while at the same time leveraging 
new geometric measurements (e.g., the ESA Gaia mission parallaxes).
In 2005 we started the SH0ES
program to use new instruments on the {\sl HST} to accomplish
this. The SH0ES ladder is composed of three components or "rungs'': Geometry to
Cepheids to Type Ia supernovae. The criticial, new elements of SH0ES are:

\begin{itemize}
  \item Cancel flux calibration errors using {\sl HST} to measure all Cepheids
  (rung 1 and 2);
  \item Observe all Cepheids in the Near-Infrared (NIR) to minimize dust;
  \item Use best quality SN data, consistently calibrated \citep[Pantheon+,][]{Brout:2022};
  \item Comprehensive error analysis, include covariance, analyze plausible 
  variants;
  \item Publicly release data, $10^7$ data numbers, and code to fit the data.
\end{itemize}

\subsection{Geometric Rung}

\begin{figure}[t!] 
\begin{center}
\includegraphics[width=0.6\textwidth]{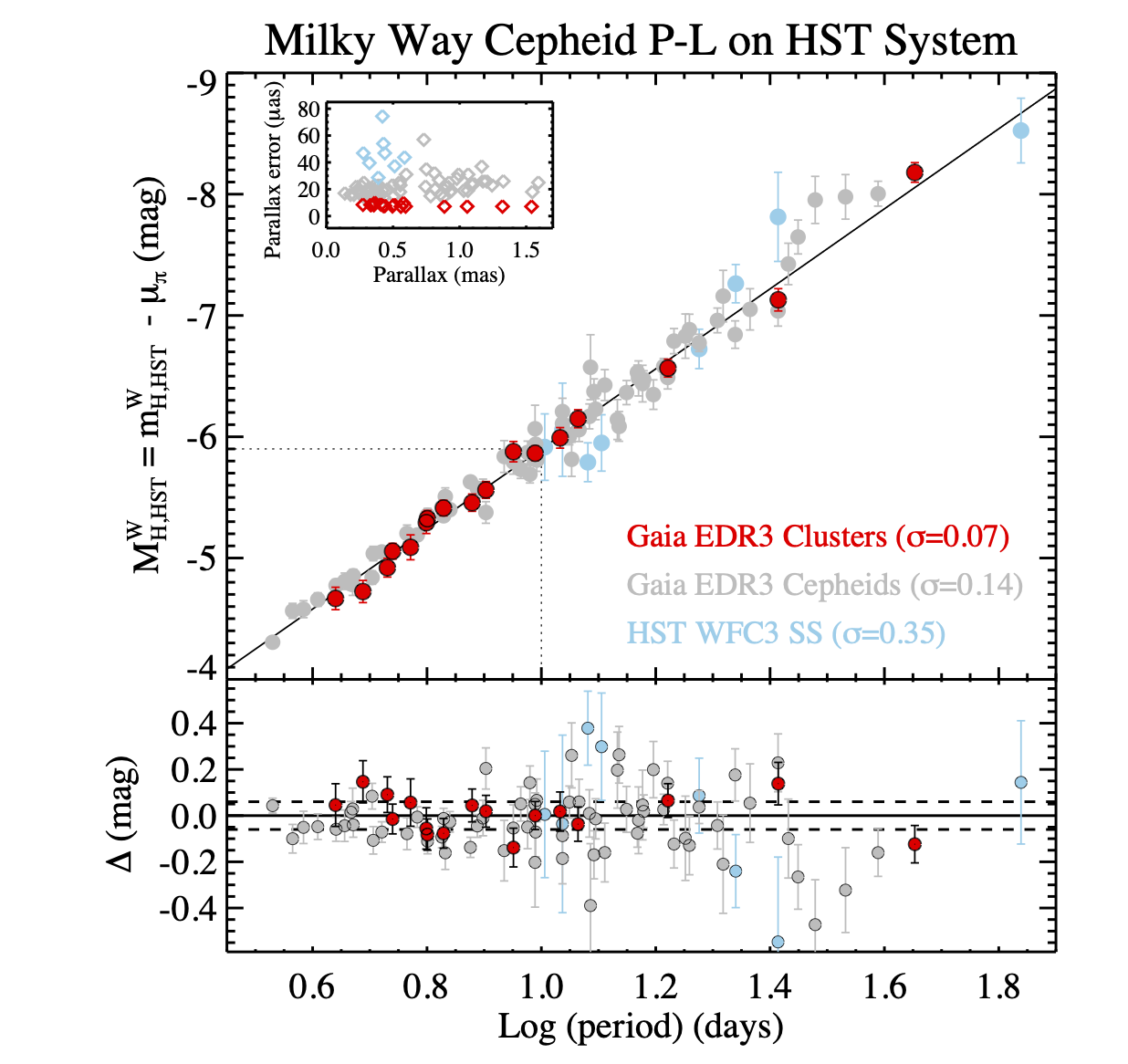}
\end{center}
\caption{The Milky Way Cepheid period--luminosity relation in the {\sl
HST} NIR, reddening-free (Wesenheit) system as calibrated with three
samples. Parallaxes from {\sl HST} spatial scanning \citep{Riess:2018} for 8
Cepheids are in blue and yield 3\% precision. The 68 points in gray
\citep{Riess:2021} using {\sl Gaia} early Data Release 3 (EDR3) parallaxes with
simultaneous calibration of the parallax offset. The red points come
from cluster Cepheids and do not require parallax offset calibration
as they are measured in the range where {\sl Gaia} is best calibrated \citep{Riess:2022clusters}.
These samples differ in their parallax precision (inset) leading to
the low dispersion of $\sigma=0.07$ mag for the cluster Cepheids.}
\label{fg:clusters} 
\end{figure}

There are three geometric anchors of the distance ladder which rely on
different systems and measurements and so are fully independent of
each other. Milky Way Cepheid parallaxes, thanks to {\sl Gaia} (now in
Data Release 3) provide $\sim$ 1\% precision in the calibration of the
Hubble constant. These are a tremendous advancement over past work.
However, we could not leverage the geometric precision these afford
without an equally precise photometric tie. The SH0ES project has
spent a lot of effort to developing a spatial scanning method for
photometric measurements of ultra bright targets with {\sl HST} that demonstrates extreme
precision and accuracy for direct calibration of Cepheids on the
second rung (in SN Ia hosts). These scans were also previously used to
measure parallaxes (before {\sl Gaia}). More recently, even greater
precision and independence from {\sl Gaia} systematics (the parallax
offset term) has come from Cepheids in open star clusters, where hundreds
of stellar parallaxes may be averaged \citep{Riess:2022clusters}.
Figure \ref{fg:clusters} shows three different approaches to measuring
Cepheid parallaxes, all in good agreement.

\begin{figure}[h] 
\begin{center}
\includegraphics[width=0.52\textwidth]{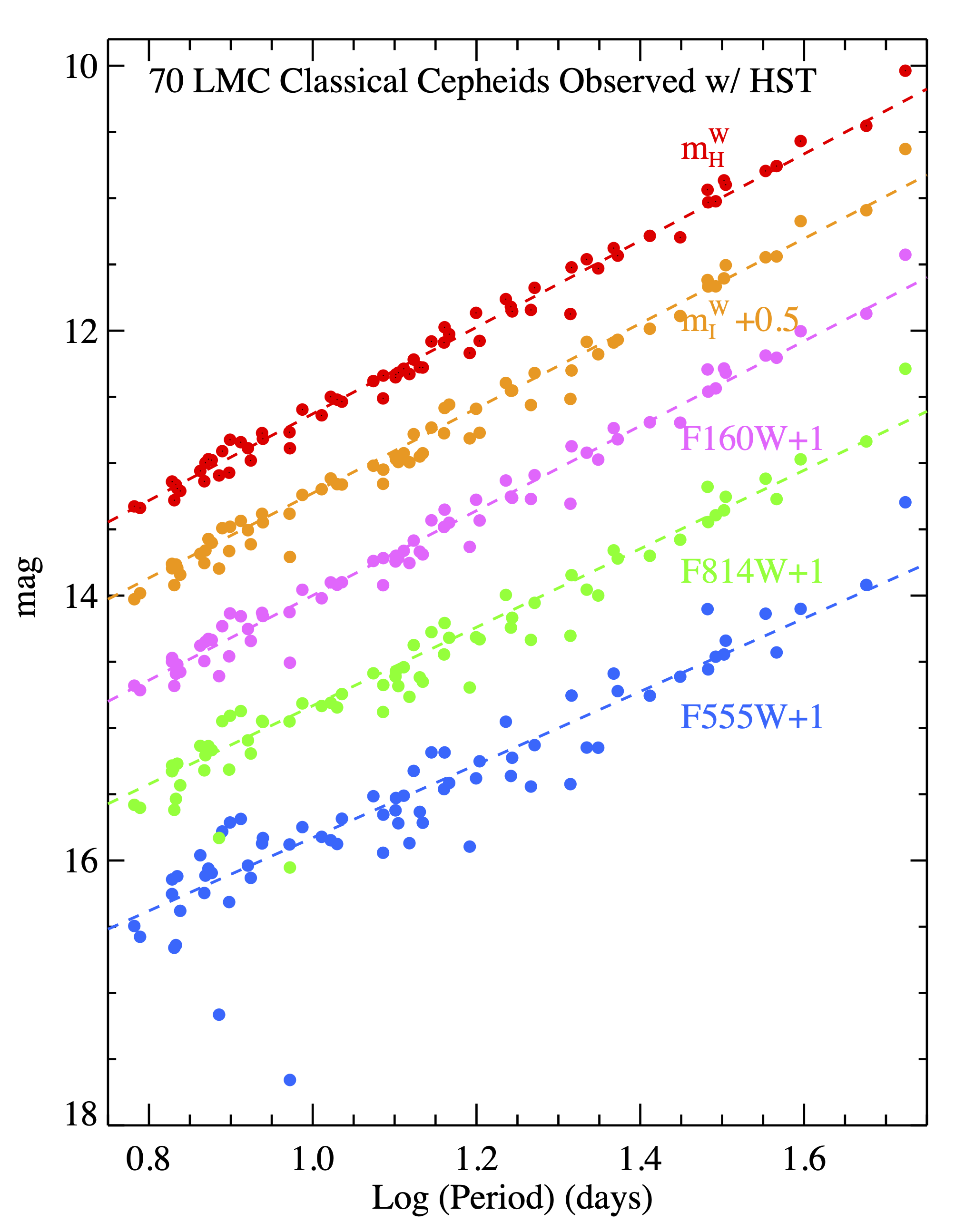}
\end{center}
\caption{Period-mean magnitude relation for the 70 LMC Cepheids observed with {\sl HST} using DASH mode.}
\label{fg:lmc} 
\end{figure}

Another anchor comes from the exquisite detached-eclipsing binary
measurements in the Large Magellanic
Cloud \citep[LMC;][]{Pietrzynski:2019} which reach 1.2\% precision.
Again, leveraging these requires direct measurements of Cepheids with
{\sl HST}. These measurements \citep{Riess:2019} yield both greater
precision and accuracy, especially in the NIR, than those from the
ground \citep[see past ground data in][in this conference]{Freedman:2023}
with a dispersion of $\sim$0.07 mag. They also demonstrate the power
of NIR data to mitigate dust and the use of colors to deredden for
remaining dust. These are shown in Figure \ref{fg:lmc}.

\begin{figure}[h] 
\begin{center}
\includegraphics[width=0.9\textwidth]{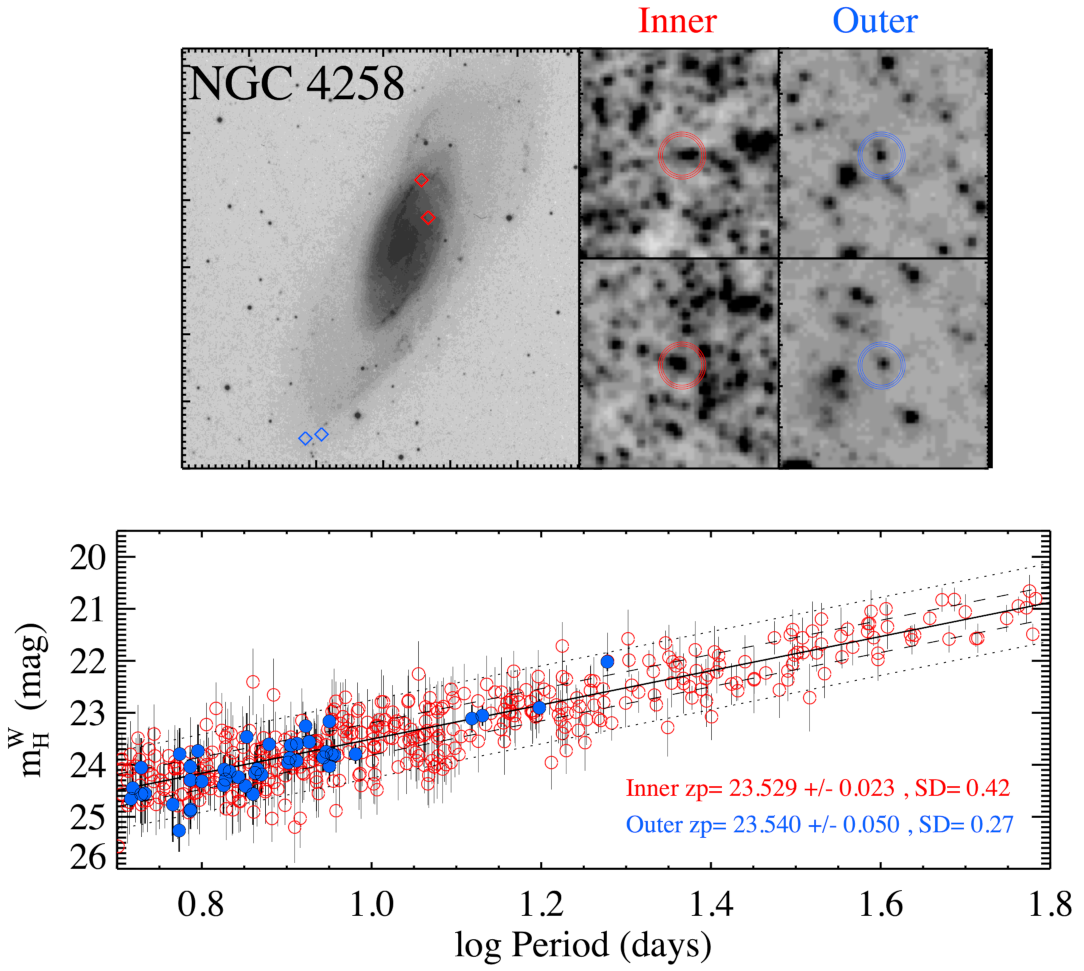}
\end{center}
\caption{Comparison of Cepheids measured in a dense (inner) field (in
red) and sparse (outer) field (in blue) of NGC 4258. Because these
Cepheids are at the same distance, the comparison shows the accuracy
of the background estimates, which differ in the mean over the same
sampled range, $0.7 < \log P < 1.2$ [days], by 0.45 mag (relative to
the Cepheids) yet yield a consistent intercept with $\sigma =0.05$
mag. The difference in metallicity between the samples, $\sim$0.08
dex, corresponds to a difference of 0.02 mag, smaller than the
precision of this comparison. }
\label{fg:ngc4258}
\end{figure}

The third anchor is the water-maser host NGC 4258, with a 1.5\%
geometric distance \citep{Reid:2019}. By collecting Cepheids in six
fields, the SH0ES Team has collected 669 Cepheids in NGC
4258 \citep{Yuan:2022}. The Cepheids in NGC 4258 also provide a strong
test of the effect of the blotchy background on Cepheid photometry,
because we can compare the Cepheid intercepts measured in the inner
region (where crowding is high) versus in the outer region, where it
is low. As shown in Figure \ref{fg:ngc4258}, the agreement is good, at
the $\sim$0.05 mag level. Here we clarify some
terminology. \citet{Freedman:2023}, at this conference, referred to
crowding as a bias in photometry. To be clear, this term (as shown in
that contribution) is the measured difference in the background level
between the true sky and the mean level of the (non-Cepheid) stars
(measured with artificial stars). Thus, the true (Cepheid) background
is both components added together. Accounting for these results in {\it no
bias}, but due to the variations in the latter, it does lead to
greater statistical noise which is not `unexplained' but, rather,
measured from the artificial stars and included in the measurement errors.
Multiple tests such as the one
above show that the photometry is accurate and provided that one has
access to more than $\sim$25 Cepheids per host, the noise does not
reach the level of the uncertainty in the SNe Ia the Cepheids
calibrate.

\subsection{SNe Ia calibrations}

The next rung in the distance ladder is the consistent measurement of
Cepheid variables in the hosts of recent SNe Ia. Cepheids can be
measured effectively with {\sl HST} up to $z \sim 0.01$ or $D < 40$
Mpc, a volume that produces a prototypical SN Ia with low reddening
about once a year. Because each SN Ia yields a distance with $\sim$6\%
precision, we need to collect many dozens in order to reach the target
goal. The SH0ES team has now collected 42 SNe Ia in 37 hosts, an
effort requiring about 1000 orbits of {\sl HST}, resulting from more
than a dozen Time Allocation Committee-selected proposals. These SNe
necessarily come from the last four decades, the era of digital
measurements and from many SNe surveys. The SNe measurements are
standardized across many surveys by the Pantheon+ collaboration using all-sky surveys.

\begin{figure}[h] 
\begin{center}
\includegraphics[width=0.85\textwidth]{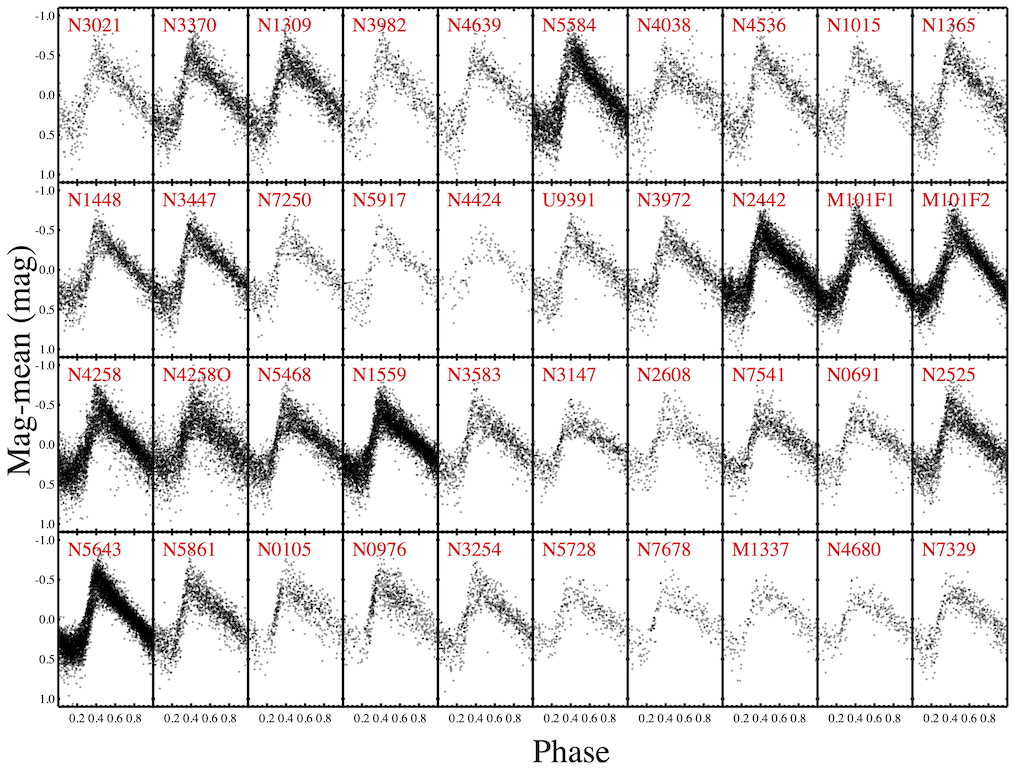}
\end{center}
\caption{Composite visual (F555W) or white-light (F350LP) Cepheid
light curves. Each {\sl HST} Cepheid light curve with $10 < P <
80$~days is plotted after subtracting the mean magnitude and
determining the phase of the observation.  }
\label{fg:lcs}   
\end{figure}

\begin{figure}[h] 
\begin{center}
\includegraphics[width=0.8\textwidth]{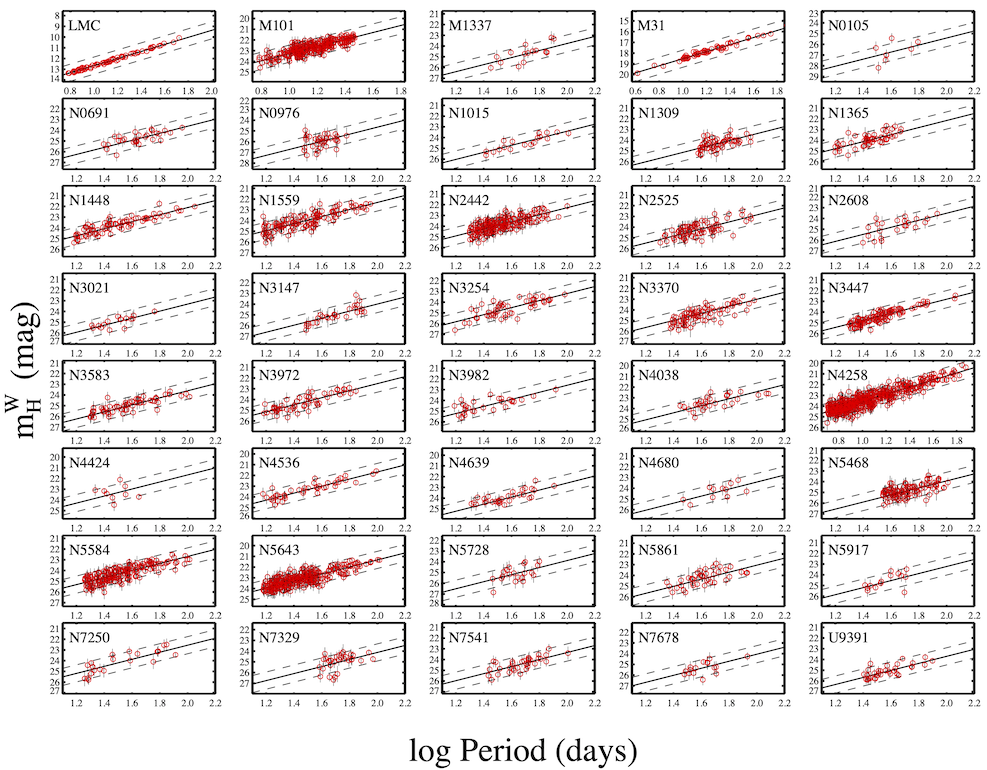}
\end{center}
\caption{{\sl HST} NIR Cepheid period--Wesenheit index relations. The
Cepheid magnitudes are shown for 37 SNe Ia hosts, M31, and two of the
three possible distance-scale anchors (LMC and NGC 4258). The
uniformity of the photometry and metallicity reduces systematic errors
along the distance ladder. A single slope is shown and used for the
baseline, but we also allow for a break (two slopes) as well as
limited period ranges in some analysis variants.}
\label{fg:pls}   
\end{figure}

The Cepheids are found in these hosts primarily with a white light
filter and then followed up in the optical ($V$ and $I$ bands) and the
NIR ($H$ band). It is important to note that the optical measurements
are not `random-phase' as claimed in the previous
contribution \citep[see][in this issue]{Freedman:2023}; they are at
known phases due determined from the simultanoues white-light light curves and,
thus, are used to measure the magnitudes at mean phase. Composite
light curves of all Cepheids in all hosts are shown in
Figure \ref{fg:lcs}. These are used to identify Cepheids and determine
their periods. The individual period--luminosity relations are shown
in Figure \ref{fg:pls}. The number of Cepheids varies but has a mean
of around 50 Cepheids per host.

\begin{figure}[h] 
\begin{center}
\includegraphics[width=0.7\textwidth]{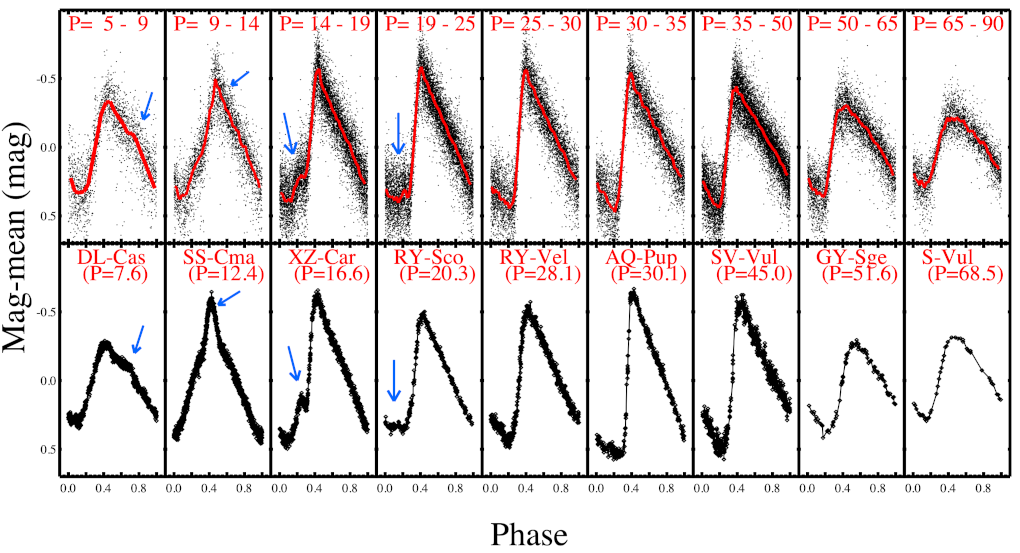}
\end{center}
\caption{Composite visual (F555W) or white-light (F350LP) Cepheid
light curves binned by period (top) and compared with individual Milky
Way Cepheids near the middle period of the bin. The `Hertzsprung
progression' (the relation between the light-curve shape and the
period) is apparent, including subtle features like the progression in
phase of a resonance `bump' between the second overtone and
fundamental pulsation for $P < 20$ days. The red line is a cubic
spline constrained by the averages of bins in phase. }
\label{fg:hr}
\end{figure}

Because this conference is focused on Cepheid variables, we can take a
closer look at the extragalactic Cepheids and address the question if
they are like those in the Milky Way. Naturally we would think so, but
we can also bin their light curves by period and compare the changing
shape to the well-known Hertzsprung progression. There is a rather
striking match, as shown in Figure \ref{fg:hr}. The light curves at
short periods are symmetric and as the periods grow, become more
asymmetric and with higher amplitudes before becoming more shallow and
sinusoidal-looking at long periods. There are also fine features like
the bump which evolves with period to be earlier in phase due to a resonance between an overtone and fundamental oscillation.

\begin{figure}[h] 
\begin{center}
\includegraphics[width=0.8\textwidth]{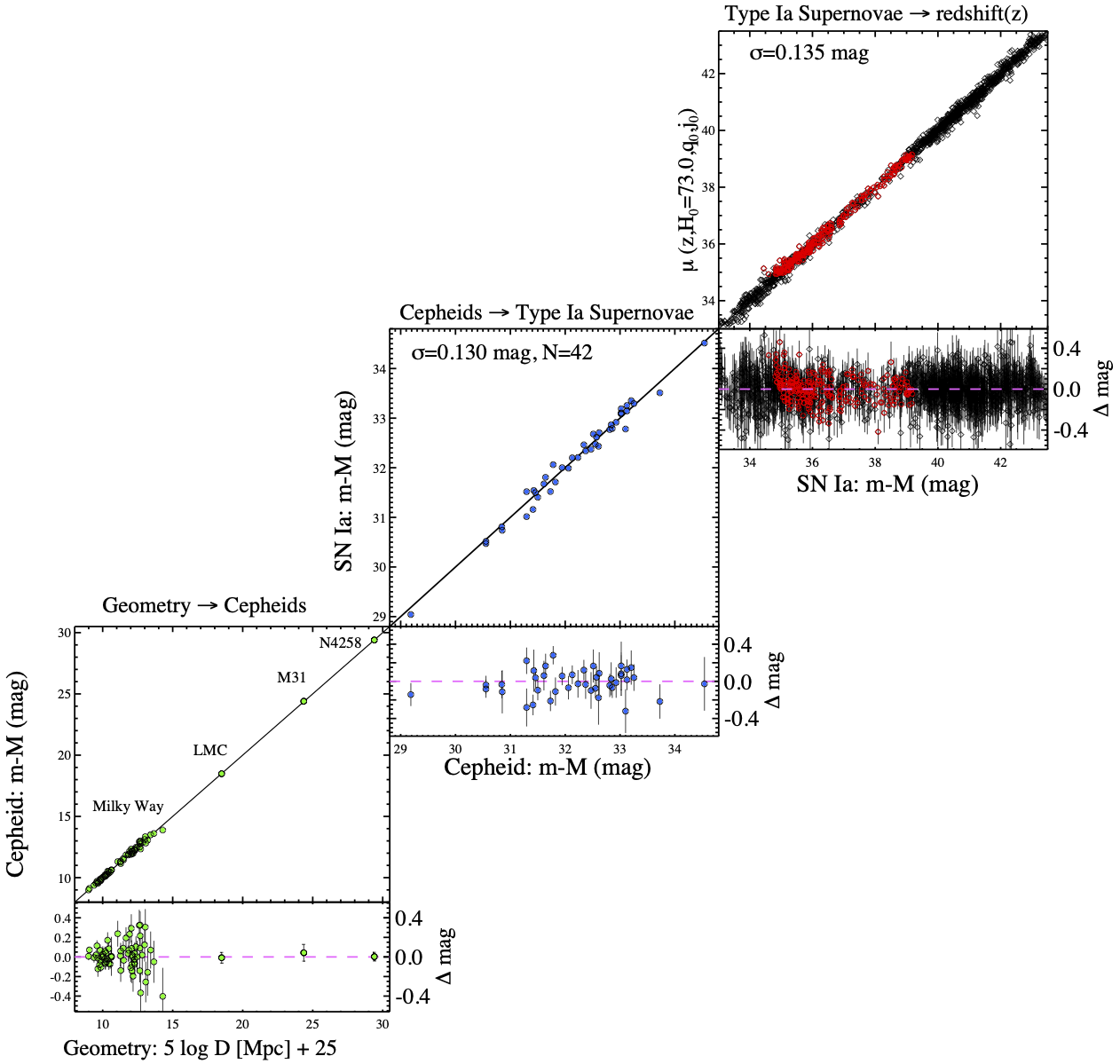}
\end{center}
\caption{Complete distance ladder \citep{Riess:2022}. The simultaneous
agreement of distance pairs: geometric and Cepheid-based (bottom
left), Cepheid- and SN-based (middle), and SN- and redshift-based (top
right) provides the measurement of H$_0$. For each step, geometric or
calibrated distances on the abscissa serve to calibrate a relative
distance indicator on the ordinate through the determination of $M_B$
or H$_0$. Results shown are an approximation to the global fit, as
discussed in the text. Red SN points are at $0.0233 < z < 0.15$, with
the lower-redshift bound producing the appearance of asymmetric
residuals when plotted against distance.}
\label{fg:dl}
\end{figure}

The final rung is the measurement of the Hubble flow from SNe Ia at
moderate redshifts. We made these measurements from $\sim$300 SNe Ia
from the Pantheon$+$ sample at $0.023 < z < 0.15$. This redshift range
was chosen to avoid the largest, local peculiar motions while avoiding
the highest redshifts to avoid senstivity to higher-order terms, $q_0$
and $j_0$. We correct for peculiar velocities using mass maps and
determine $q_0=-0.55$ ($j_0=1$) from high-redshift SNe Ia. The global fit of the
distance ladder and H$_0$ is a large set of simultaneous, linear
equations which can be optimized for five global parameters, including
the fiducial luminosity of SNe Ia and Cepheids, two parameters
specific to Cepheids, the slope of their period--luminosity relation
and metallicity dependence, and the final parameter is (5 log
H$_0$). The entire data set as fitted is shown in
Figure \ref{fg:dl}. The equations can be solved exactly by inverting a
matrix, but we also use a brute-force, Markov Chain Monte Carlo
approach to view the interdependency of parameters.

\begin{figure}[h] 
\begin{center}
\includegraphics[width=0.7\textwidth]{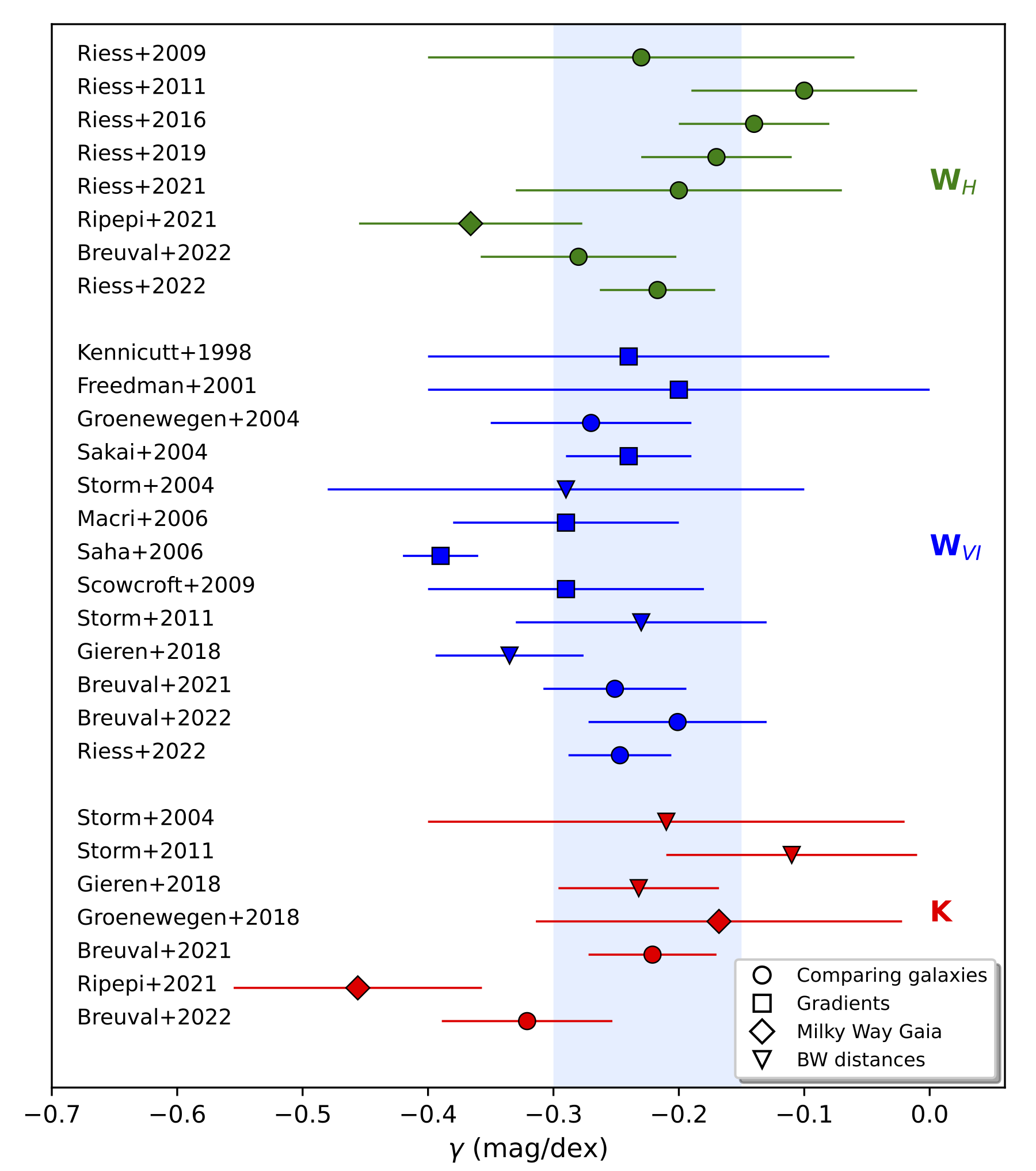}
\end{center}
\caption{Recent Cepheid metallicity measurements.  There is good consistency on the sign (negative) and scale, -0.1 to -0.3 mag dex$^{-1}$, for this term.  It is important to know that this term has little effect on $H_0$ because the mean metallicity of Cepheids on the first and second rung is similar.}
\label{fg:metallicity}
\end{figure}

One of the more interesting parameters for this group is the Cepheid
metallicity dependence. We find a value of $-0.22 \pm 0.04$ mag
dex$^{-1}$, which sits well in the middle range of recent findings.
Figure \ref{fg:metallicity} shows a summary of metallicity results: we 
see that the metallicity effect ($\gamma$) has now converged to a 
modest range of values with wide concurrance on a negative sign, with a 
good agreement between various measurements from independent teams and 
methods. Although a few studies diverge from this central value, 
\citet{Breuval:2022} identified that these measures where affected by 
geometry effects in the Magellanic Clouds and by the use of an 
atypical {\em Gaia} parallax zero-point in the Milky Way. Overall 
we find strong evidence, from dozens of calibrations in the literature, 
that metal-rich Cepheids are brighter than metal-poor ones by  
$\sim$0.2 mag dex$^{-1}$. There is also agreement on the size and sign between 
the empirical measurements and theoretical modeling (presented at this meeting) 
of Wesenheit magnitudes (the only dissagreement with the theory was on the 
monochromatic dependence) with Wesenheit magnitudes of greatest relevance 
for distance measurements. However, it is important to note that the value of 
the Hubble constant is not sensitive to this parameter, because the mean 
metallicity of the SNe Ia hosts is similar to the mean of the anchors. The 
anchors themselves have somewhat different metallicities, so the metallicity
term is an important contributor to the consistency of the anchors.

The resulting determination of H$_0$ is $73.04 \pm1.04$ km s$^{-1}$
Mpc$^{-1}$, in good agreement with the mean of local determinations
but inconsistent with the prediction of $67.4 \pm 0.5$ km s$^{-1}$
Mpc$^{-1}$ at the 5$\sigma$ confidence level. This discrepancy is so
famous, it has a name, the `Hubble Tension', and it has been
disturbing the cosmology community for a decade. It is certainly
tempting to think the discrepancy may be telling us we are missing
something in the cosmological model.

We also calculated 67 variants of the analysis varying many, many,
many things. There are too many to discuss them all, but we select a
subset of the most interesting and what value of H$_0$ they yield. \\

{\bf Variants}
\begin{itemize}
\item Optical Cepheid data only (72.7)
\item Different peculiar velocity map or none (73.1, 72.7)
\item SN scatter monochromatic + mass step (73.5) 
\item No pre-2000 SNe (73.2) 
\item Closest half hosts (73.1)
\item Most crowded half (73.4)
\item Least crowded half (73.3)
\item Skip `local hole' $z>0.06$ (73.4)
\item All host types (73.3)
\item Include tip of the red-giant branch (consistent) jointly (72.5)
\item No metallicity term (73.5)
\item Break in the period--luminosity relation at $P=10$ days (72.7)
\item No extinction correction (74.8)
\item Individual host extinction law (73.9)
\item Free parametric extinction law (73.3)
\item Low $R_V =2.5$ extinction law (73.2)
\item Two of three anchors (73.0, 73.4, 73.2)
\item No outlier rejection (73.4) \\
\end{itemize}

\begin{figure}[h] 
\begin{center}
\includegraphics[width=0.85\textwidth]{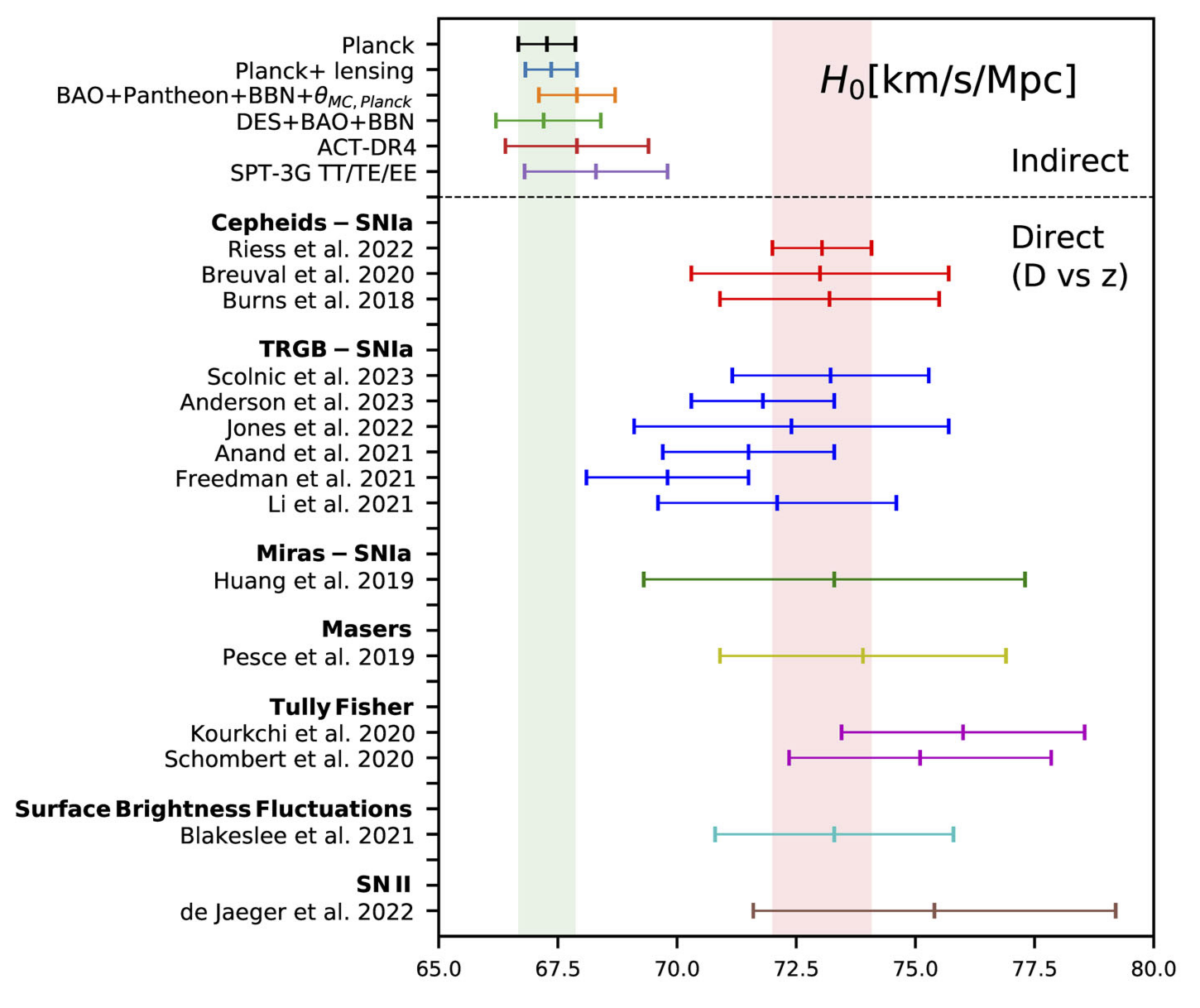}
\end{center}
\caption{Recent measures and the Hubble Tension. The subset of recent
measures was selected to be those which are the most cited, offer the
most independence, and use the best source of data when multiple are
available (e.g., {\sl HST} versus ground). Figure from \citet{DiValentino:2021}.}
\label{fg:tension}
\end{figure}

As shown, it is hard to move the value of the Hubble constant much
below 72.5 or much above 73.5, which indicates that the baseline
result is fairly robust. Although the SH0ES team measurement provides
the highest precision, a consequence of having the largest sample of
calibrated SNe Ia, the Hubble Tension is a broad finding that has
lasted a decade and covers a wide range of different techniques, teams
and methods. A recent summary of many measures is shown in
Figure \ref{fg:tension}. This is a subset of recent measures but
selected to be those which are the most cited, offer the most
independence, and use the best source of data when multiple are
available (e.g., {\sl HST} versus ground).

\section{What can we learn from {\sl JWST}?}

There are many tests of the distance ladder to discuss, but in the
interest of space, we will focus on the most recent and novel from the
{\sl James Webb Space Telescope} ({\sl JWST}). Observations with the
{\sl JWST} offer for the first time the resolution in the NIR (and
mid-IR) to separate Cepheids from their stellar crowds. This greatly
reduces the noise in the period--luminosity relations, whose source
was the variations in background (i.e., primarily fluctuations in the density of
red giants) seen at {\sl HST} NIR resolution. The previous contribution
\citep[see][this volume]{Freedman:2023} presented their group's
{\sl JWST} observations following up Cepheids discovered with {\sl
HST} in the SN Ia host NGC 7250. They showed 14 Cepheids with a
dispersion of 0.22 mag at 1.15 $\mu$m, half the size of the dispersion
seen with {\sl HST} in the NIR (see also the wonderful contribution by
Kayla Owens presenting this work). This is consistent with what we see
in our {\sl JWST} observations.

We observed $>$320 Cepheids in two hosts, NGC 5584 (the host of a SN
Ia) and NGC 4258, the previously discussed calibrator. We observed the
Cepheids in three filters centered at 0.9, 1.5 and 2.8 $\mu$m and
across two epochs separated by 2--3 weeks. The differences in epochs
allowed us to (1) confirm the astrometric identification of the
Cepheids through their variability; (2) check the photometric
stability of {\sl JWST} over time by comparing non-variable stars; and
(3) constrain the phase of the Cepheids (knowledge of which has
dissipated over the years since they were discovered).

\begin{figure}[h] 
\begin{center}
\includegraphics[width=0.98\textwidth]{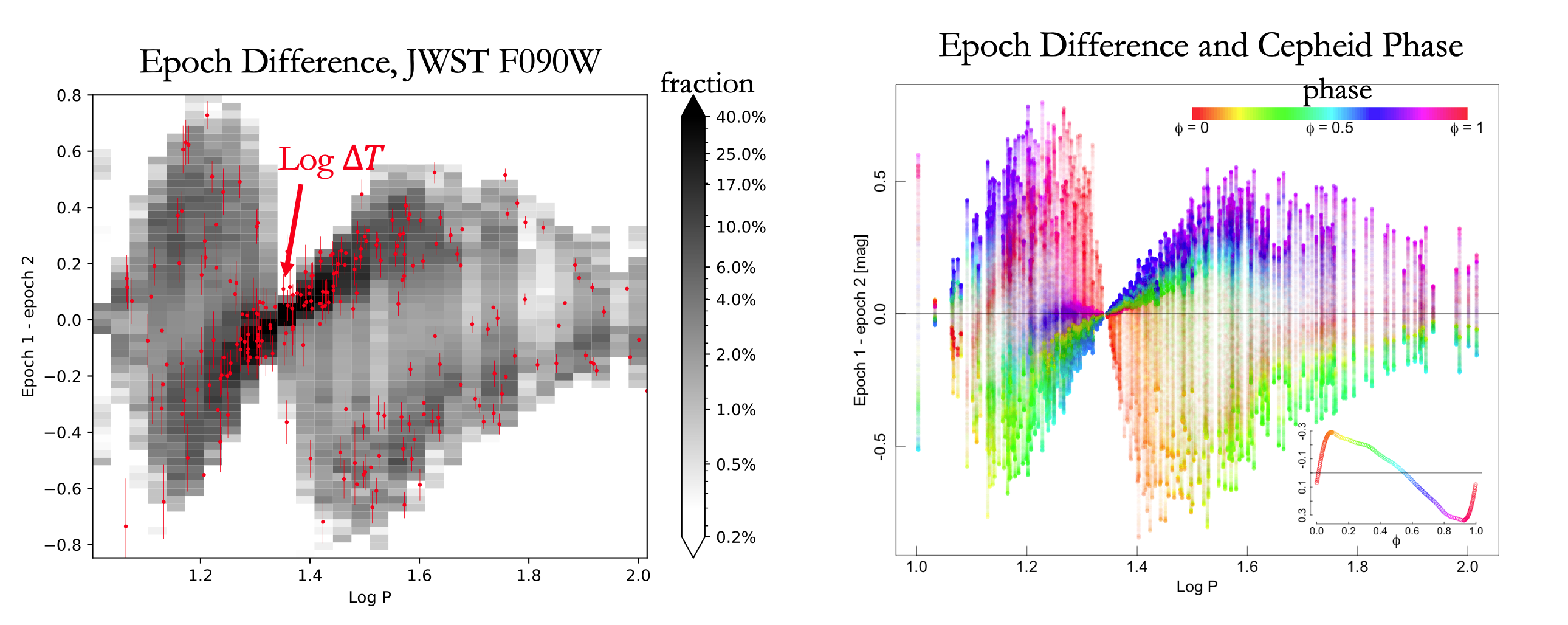}
\end{center}
\caption{{\sl JWST} two-epoch measurements of Cepheids in NGC
5584. (left) Epoch differences. (right) The expected differences which
determine the phase.}
\label{fg:phaseing}
\end{figure}

\begin{figure}[h] 
\begin{center}
\includegraphics[width=0.75\textwidth]{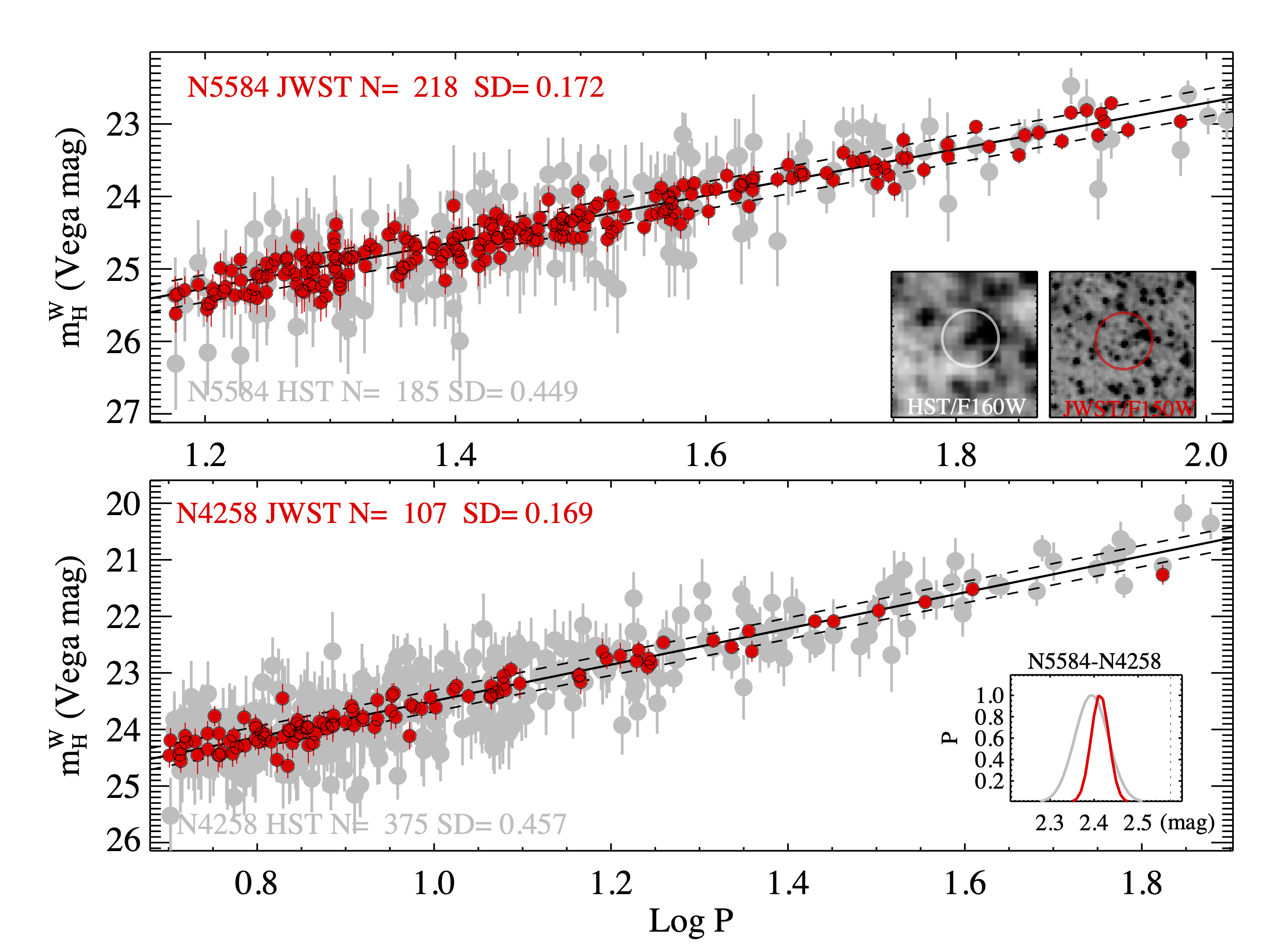}
\end{center}
\caption{Comparison with the standard \citep[SH0ES;][]{Riess:2022}
magnitude $W^H_{V,I}$ period--Wesenheit relation used to measure
distances. The red points use {\sl JWST} F150W, and the gray points
are from {\sl HST} F160W, including a small transformation,
F150W$-$F160W $= 0.033+0.036[(V-I)-1.0]$. The top panel is for NGC
5584, with the inset showing image stamps of the same Cepheid seen in
the $H$ band by each telescope. The bottom panel is for NGC 4258, with
the inset showing the difference in distance moduli between NGC 5584
and NGC 4258 as measured with each telescope.}
\label{fg:R23_mp} 
\end{figure}

\begin{figure}[h] 
\begin{center}
\includegraphics[width=0.73\textwidth]{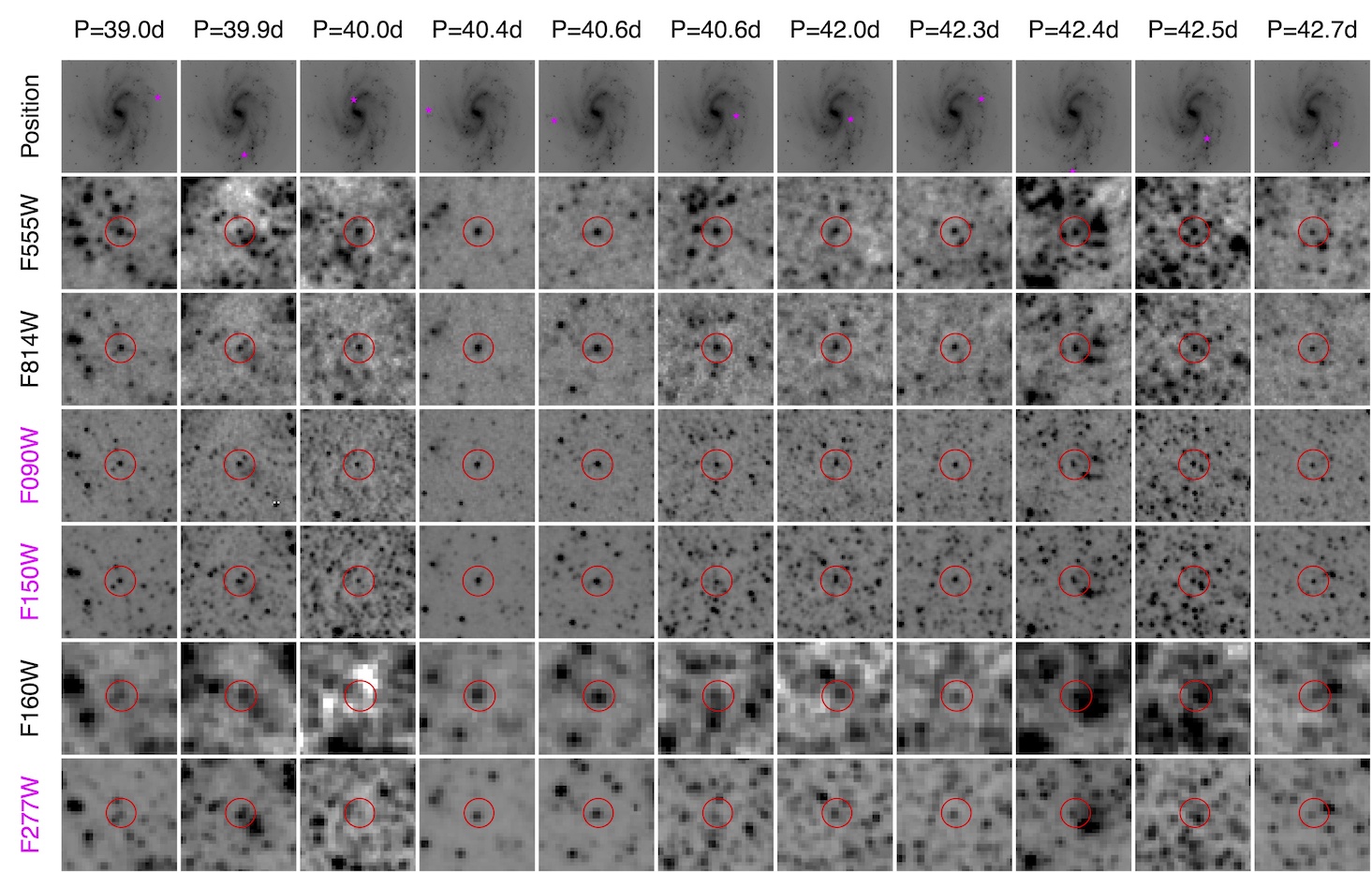}
\end{center}
\caption{{\sl HST} and {\sl JWST} image stamps of NGC 4258 for all
Cepheids with periods of 18--41 days. The top row indicates the
location of each Cepheid. {\sl HST} filters are labeled in black and
{\sl JWST} filters in magenta. }
\label{fg:stamps}  
\end{figure}

In Figure \ref{fg:phaseing} we show the use of the epoch differences
applied to the Cepheid phases. We found that the combination of two
epochs (to reduce the random phase error) and the ability to constrain
the phase reduced the dispersion from $\sim$0.43 mag to $\sim$0.17
mag, which is equivalent in statistical leverage to doubling the sample
size. The results for both hosts and direct comparison with {\sl HST}
is shown in Figure \ref{fg:R23_mp}. In Figure \ref{fg:stamps} we show
a comparison of many postage stamps around Cepheids with {\sl HST} and
{\sl JWST}. The contrast offered by {\sl HST} is excellent in the
optical, but becomes low, and the crowding high, in the NIR. This is
the region where {\sl JWST} wins by a large margin. The filters F160W
for {\sl HST} and F150W for {\sl JWST} are nearly the same, which
offers the best comparison.

Because we have measured Cepheids with {\sl JWST} at both rungs, an SN
Ia host and NGC 4258, we can directly measure and compare the
resulting calibration they provide for the distance ladder, a
difference which is independent of uncertainties in the zero points of
{\sl JWST}, which cancel out. As shown in the inset of
Figure \ref{fg:R23_mp}, they provide a consistent difference (in
magnitudes) between rungs as seen with {\sl HST}. Because the Cepheid
samples are large ($>$320 with {\sl JWST} compared with $>$560 with
{\sl HST}) and this is purely a test of Cepheid measurements, the
precision of this test is very high, $0.04$ mag, many times
smaller than the 0.18 mag size of the Hubble Tension. So, by stripping the noise
and crowding from the Cepheid measurements with {\sl HST}, these
observations provide the strongest evidence yet that systematic errors
in {\sl HST} Cepheid photometry do not play a significant role in the
present Hubble Tension \citep{Riess:2023}. However, a more ambitious
goal which will require more observations and the work of many is to
improve the precision of the local measurement of H$_0$ to reach
1.0\%.

So, what causes the Hubble Tension? We do not know. The facts are that
it has lasted 10 years, it is statistically very significant and all
precise (e.g., $\leq$3\%) local ladder measurements are higher than
the CMB-based prediction, so it is pretty robust. We have undertaken a
rather exhaustive set of tests and studies of systematics and there
has been no indication of a problem on the measurement side. In this
regard the {\sl JWST} data are very impressive. One can speculate
about an unknown systematic error, but the data have become very good,
so concrete proposals for such systematics are inconsistent to date
with the breadth of data and tests. It is also easy to speculate that
it is something missing in $\Lambda$CDM but much harder to come up
with a specific notion of what that is with sufficient rigor to test
and yield a better fit to the cosmological data. This is a very hot
topic in the theory community and we are seeing lots and lots of
proposals, but nothing definitive or compelling (yet). We are
optimistic. History has shown that basic errors are found by our
community rapidly (timescale of $\sim$ months) and that discrepancies that last this long
are usually telling us something interesting and thus worth the investment.
Let's be curious.

\begin{figure}[h] 
\begin{center}
\includegraphics[width=0.8\textwidth]{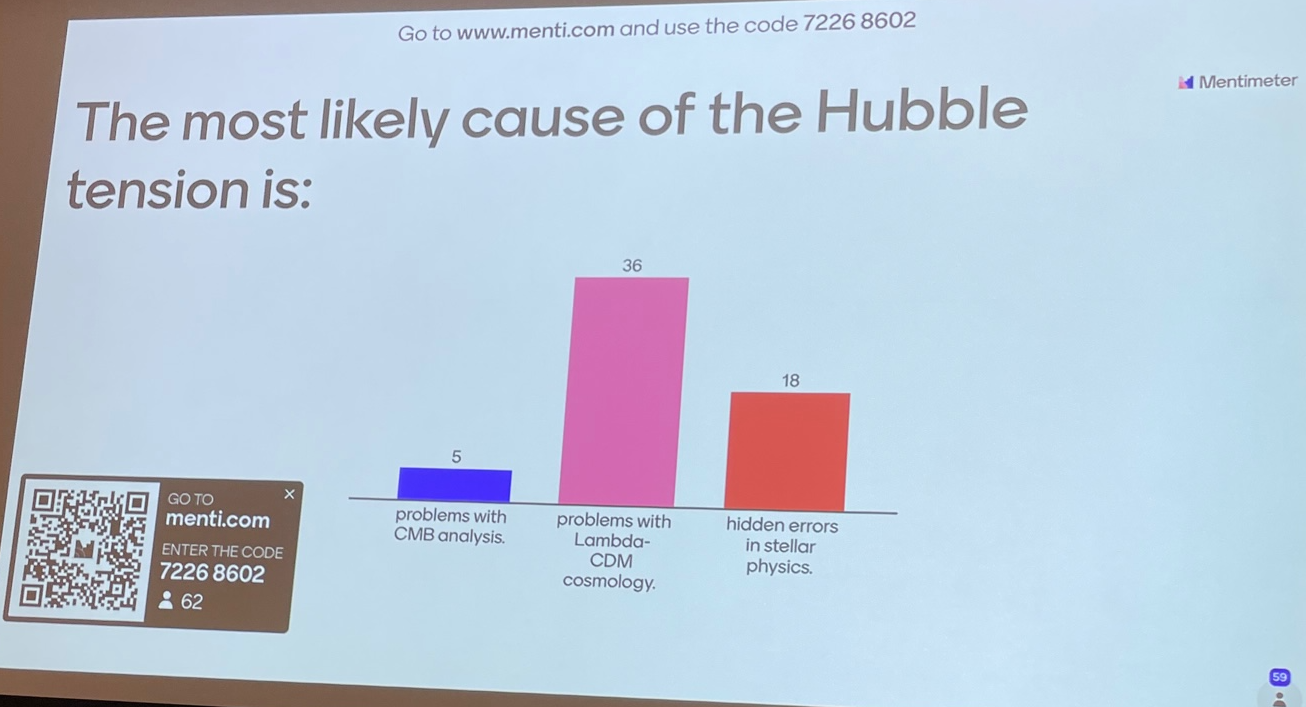}
\end{center}
\caption{Meeting Poll. }
\label{fg:poll} 
\end{figure}

(On another day the conference enacted a poll, which is a rather
unscientific way to resolve the Hubble Tension but it is a snapshot of
what people at the conference thought after a couple of days, and
entertaining, and we attach the result as Figure \ref{fg:poll}. The
polling was anonymous and used a smartphone app to allow the
participants to vote. The three options voted on as possible cause of
the Hubble Tension were a problem with the CMB analysis, a problem
with $\Lambda$CDM or an issue with stellar physics. As the
conference was a gathering of people studying stars and their
physics, it was noteworthy that $\sim$70\% of attendees did not think
the Hubble Tension was due to a problem with the physics of stars). \\

We want to thank the organizers for a wonderful conference filled with
discussions and goulash.

\appendix
\section{From the CMB to H$_0$, a primer}

The first and strongest peak we see in the power spectrum (expansion
into spherical harmonics), $l_{\rm s}$, determines the angle subtended
by the sound horizon, the distance a fluctuation in the plasma will
travel at the sound speed from the time of the Big Bang to the surface
of last scattering (i.e., $z\sim1000$). This scale results from the
angular size of this feature, $\ell_{\rm s} \simeq 2/\theta_{\rm
s}$. The angle subtended by the sound horizon is $\theta_{\rm s}=
r_{\rm s}/D_A$, where $D_A$ is the angular-diameter distance to the
CMB surface of last scatter, and $r_{\rm s} \sim c_{\rm s} t_{\rm
dec}$ is the sound horizon. The angle is exceedingly well-measured by
{\sl Planck}: $\theta_{\rm s}=(1.04109\pm 0.00030)\times 10^{-2}$ rad. 

We can calculate the physical size of the sound horizon from the
physics of the early Universe as given by $\Lambda$CDM. It follows by
integrating the sound speed $c_{\rm s}(t)$ over time prior to recombination
and is given by the integral,
\begin{equation}
    r_{\rm s} = \int_{z_{\rm ls}}^\infty \frac{c_{\rm s}(z)\, {\rm
    d}z}{H(z)} = \frac{c}{\sqrt{3}H_{\rm ls}} \int_{z_{\rm
    ls}}^\infty \frac{{\rm d}z} {\left[\rho(z)/\rho(z_{\rm
    ls}) \right]^{1/2} \left(1 + R \right)^{1/2}},
\end{equation}
over redshift $z$. Here, $z_{\rm ls}\simeq1080$ is the redshift at
which CMB photons last scatter, $c_{\rm s}(z) = c \left[
3(1+R) \right]^{-1/2}$ is the sound speed of the photon--baryon fluid,
with $R=(3/4)(\omega_{\rm b}/\omega_\gamma)/(1+z)$, and $\rho(z)$ is
the total energy density at redshift $z$. Here, $\omega_{\rm
b}= \Omega_{\rm b} h^2$ is the current physical baryon density, where
$h\equiv {\rm H}_0/(100\, {\rm km}\, {\rm s}^{-1}\, {\rm Mpc}^{-1})$
is a dimensionless Hubble constant. $\omega_{\rm b}$ is a parameter
constrained by higher-peak features in the CMB power spectrum (Silk
damping at higher $l$ and the relative heights of the even- and
odd-numbered peaks with $\omega_{\rm m}$) and $\omega_\gamma
=2.47 \times 10^{-5}$ is the physical photon energy density given by
COBE FIRAS. It is convenient to define the expansion rate at last
scattering,
\begin{equation}
    H_{\rm ls} = 100\, {\rm km}\, {\rm s}^{-1}\, {\rm
    Mpc}^{-1}\, \omega_{\rm r}^{1/2} (1+z_{\rm ls})^2 \sqrt{
    1+ \frac{\omega_{\rm m}}{\omega_{\rm r}}\frac{1}{1+z_{\rm ls}} },
\label{eqn:Hls}    
\end{equation}
where $\omega_{\rm m}=\Omega_m h^2$ is the present phsyical matter
density fit from the higher-peak structure in the CMB. In the standard
cosmological model, the early-Universe energy density is
$\rho(z) \propto \omega_{\rm m} (1+z)^3 + \omega_{\rm r}(1+z)^4$. The
physical radiation density is
\begin{equation}
     \omega_{\rm r} = \left[ 1 + \frac78 N_{\rm
     eff} \left( \frac{4}{11} \right)^{4/3} \right]\omega_\gamma,
\end{equation}
where the second term accounts for additional non-relativistic degrees
of freedom. In the standard cosmological model, these include the
three neutrino mass eigenstates, and $N_{\rm eff}=3.06$ .

The (comoving) angular-diameter distance to the surface of last
scattering is then an integral,
\begin{equation}
    D_A = \frac{c}{{\rm H}_0} \int_{0}^{z_{\rm ls}} \frac{{\rm
    d}z}{ \left[\rho(z)/\rho_0 \right]^{1/2}},
\label{eqn:angulardiameterdistance}
\end{equation}
from recombination until the current time $t_0$, when the total energy
density is $\rho_0$. The denominator here is $\rho(z)/\rho_0
= \Omega_{\rm m}(1+z)^3+(1-\Omega_{\rm m})(1+z)^{-3(1+w)}$ in the
standard cosmological model, with a dark-energy equation-of-state
parameter $w$. The cosmological constant corresponds to $w=-1$.

From $\theta_{\rm s} = r_{\rm s}/D_A$, we infer a Hubble constant,
\begin{equation}
     {\rm H}_0 = \sqrt{3} H_{\rm ls} \theta_{\rm
    s} \frac{ \int_{0}^{z_{\rm ls}} {\rm
    d}z\, \left[\rho(z)/\rho_0 \right]^{-1/2} } { \int_{z_{\rm
    ls}}^\infty {\rm d}z\, \left[\rho(z)/\rho(z_{\rm
    ls}) \right]^{-1/2} \left(1 + R \right)^{-1/2} }
\label{eqn:cmbH0}
\end{equation}
from the CMB.  

By this route, H$_0$ depends on $\omega_{\rm b}$, $\omega_{\rm m}$,
$\omega_\gamma$, $N_{\rm eff}$ and also (to a lesser degree) on the
other model parameters through degeneracies in the description of the
fluctuation spectrum.


\begin{thebibliography}{}

\bibitem[Breuval et al.(2022)]{Breuval:2022} Breuval, L., Riess, A. G., Kervella, P. et al., 2022, ApJ 939, 89
\bibitem[Brout et al.(2022)]{Brout:2022} Brout, D., Taylor, G., Scolnic, D., et al., 2022, ApJ 938, 111
\bibitem[Di Valentino (2021)]{DiValentino:2021} Di Valentino, E. , 2021, MNRAS 502, 2065
\bibitem[Freedman and Madore(2023)]{Freedman:2023} Freedman, W. L. \& Madore B. F., 2023, arXiv:2308.02474
\bibitem[Kamionkowski and Riess(2022)]{Kamionkowski:2023} Kamionkowski, M. \& Riess, A. G., 2022, arXiv:2211.04492
\bibitem[Pietrzy\'nski et al.(2019)]{Pietrzynski:2019} Pietrzy\'nski, G., Graczyk, D., Gallenne, A., et al., 2019, Nature 567, 200
\bibitem[Planck Collaboration et al.(2020)]{Planck:2018} Planck Collaboration, Aghanim, N., et al., 2020, A\&A 641, A6
\bibitem[Reid et al.(2019)]{Reid:2019} Reid, M. J., Pesce, D. W. and Riess, A. G., 2019, ApJL 886, L27
\bibitem[Riess et al.(2018)]{Riess:2018} Riess, A. G., Casertano, S., Yuan, W., et al., 2021, ApJ 855, 136
\bibitem[Riess et al.(2019)]{Riess:2019} Riess, A. G., Casertano, S., Yuan, W., et al., 2019, ApJ 876, 85
\bibitem[Riess et al.(2021)]{Riess:2021} Riess, A. G., Casertano, S., Yuan, W., et al., 2021, ApJL 908, L8
\bibitem[Riess et al.(2022a)]{Riess:2022clusters} Riess, A. G., Breuval, L., Yuan, W., et al., 2022, ApJ 938, 36
\bibitem[Riess et al.(2022b)]{Riess:2022} Riess, A. G., Yuan, W., Macri, L. M., et al., 2022, ApJL 934, L7
\bibitem[Riess et al.(2023)]{Riess:2023} Riess, A. G., Anand, G. S., Yuan, W., et al., 2023, arXiv:2307.15806
\bibitem[Sandage et al.(2006)]{Sandage:2006} Sandage, A., Tammann, G. A., Saha, A., et al., 2006, ApJ 653, 843 
\bibitem[Yuan et al.(2022)]{Yuan:2022} Yuan, W., Macri, L. M., Riess, A. G. et al., 2022, ApJ 940, 64

\end{thebibliography}
\end{document}